\documentclass[twocolumn,superscriptaddress,preprintnumbers,amsmath,amssymb,floatfix]{revtex4-1}
\pdfoutput=1
\usepackage{graphicx} 
\usepackage{dcolumn} 
\usepackage{bm}
\usepackage{color}
\usepackage{hyperref}
\usepackage{graphicx}
\usepackage{amsmath}
\usepackage{setspace}
\usepackage{appendix}

\begin{document}
\title{Exploring the nuclear momentum anisotropy based on intermediate-energy heavy-ion collisions}

\author{Xiao-Hua Fan}
\affiliation{School of Physical Science and Technology, Southwest University, Chongqing 400715, China}
\affiliation{RIKEN Nishina Center, Wako, Saitama 351-0198, Japan}
\affiliation{Department of Physics, Graduate School of Science, The University of Tokyo, Tokyo 113-0033, Japan}

\author{Zu-Xing Yang}
\email{zuxing.yang@riken.jp}
\affiliation{RIKEN Nishina Center, Wako, Saitama 351-0198, Japan}
\affiliation{School of Physical Science and Technology, Southwest University, Chongqing 400715, China}

\author{Peng-Hui Chen}
\affiliation{College of Physics Science and Technology, Yangzhou University, Yangzhou, Jiangsu 225002, China}

\author{Zhi-Pan Li}
\affiliation{School of Physical Science and Technology, Southwest University, Chongqing 400715, China}

\author{Wei Zuo}
\affiliation{Institute of Modern Physics, Chinese Academy of Sciences, Lanzhou 730000, China}
\affiliation{School of Nuclear Science and Technology, University of Chinese Academy of Sciences, Beijing 100049, China}

\author{Masaaki Kimura}
\affiliation{RIKEN Nishina Center, Wako, Saitama 351-0198, Japan}

\author{Shunji Nishimura}
\affiliation{RIKEN Nishina Center, Wako, Saitama 351-0198, Japan}

\begin{abstract}

We simulate ultra-central collisions of prolate uranium-uranium nuclei at intermediate energies using the isospin-dependent Boltzmann-Uehling-Uhlenbeck model to investigate the impact of momentum anisotropy on spatial geometric effects. 
By defining the quadrupole deformation parameter in momentum space $\beta_\text{p}$, we establish an ellipsoidal Fermi surface, aligning its rotational symmetry axis with the one in coordinate space.
It is found that oblate momentum density enhances elliptic flow $v_2$, while prolate momentum density has the opposite effect, particularly pronounced in the outer, high transverse momentum $p_\text{t}$ region. 
Momentum anisotropy also causes differences in the initial momentum mean projection along the beam direction, with larger projections producing more pion mesons. 
Additionally, significant effects on mean square elliptic flow are observed in non-polarized collisions.
We further examine the relationship between the $v_2$-$p_\text{t}$ slope and $\beta_\text{p}$, eliminating systematic errors through the two-system ratio. 
These findings provide important references for experimentalists in heavy-ion collisions and valuable feedback to theorists regarding nuclear structure.
\end{abstract}

\maketitle

\section{Introduction}

The shape of atomic nuclei has long attracted widespread attention from nuclear physicists \cite{Bohr1970PhysicsToday23.5860, Ring1980., Moeller2016At.DataNucl.DataTables109110.1204, Heyde2011Rev.Mod.Phys.83.14671521}.
Traditionally, collective motion leads to characteristic rotational spectra in nuclear-excited states.
The electric multipole transition probability $B(En)$ between low-lying rotational states with an angular momentum difference of $n\hslash$ can be utilized to deduce the shape parameters, which in turn helps in understanding various shape-related phenomena \cite{Heyde2011Rev.Mod.Phys.83.14671521, Togashi2016Phys.Rev.Lett.117.172502, Heyde2016Phys.Scr.91.083008, Frauendorf2018Phys.Scr.93.043003, Zhou2016Phys.Scr.91.063008}.
Since 2000, researchers have developed a method based on relativistic heavy-ion collisions to study nuclear deformations in coordinate space \cite{Li2000Phys.Rev.C61.021903, Zhang2022Phys.Rev.Lett.128.022301, Jia2022Phys.Rev.C105.014905}, allowing for a more intuitive exploration of nuclear shapes.
With the discovery of linear response relation, i.e., the collective flow $v_n$ after collisions is proportional to the ellipticity $\epsilon_n$ of the overlapping region in the transverse plane of the initial reaction nuclei, it is only a matter of time before precise measurements of shape information for stable nuclei with limited shape fluctuations become achievable.

However, to date, there has been limited reaction research on the presence of deformation in the momentum space of atomic nuclei.
Previous studies on nuclear deformation via heavy-ion collisions have typically employed reaction energies of several hundred GeV/nucleon \cite{Zhang2022Phys.Rev.Lett.128.022301, Jia2022Phys.Rev.C105.014905}, where the high beam energy generally overshadows the effects of initial momentum.
Recently, research on nuclear deformation in heavy-ion collisions has been theoretically extended to the intermediate-energy regime \cite{Fan2023Phys.Rev.C108.034607, Yang2024Phys.Lett.B848.138359}, necessitating a thorough understanding of anisotropy in momentum space.
Anisotropy in momentum space is supported by the two points: 
1. In traditional nuclear structure studies,  wave functions in coordinate space and momentum space can be obtained through mutual Fourier transforms. 
This implies that the momentum distribution of a deformed nucleus should also reflect deformation.
2. Recent experiments \cite{2018Nature560.617621} indicate that the short-range correlations (SRC) lead to protons and neutrons exhibiting similar high-momentum tails in momentum space. 
However, the Fermi surfaces of protons and neutrons differ, necessitating the introduction of discordant concepts, such as the momentum correction factor \cite{Fan2022Phys.Lett.B834.137482} and the momentum gap \cite{Yong2018Phys.Lett.B776.447450}, to interpret the SRC is dominated by $np$ correlations.
Incorporating momentum anisotropy would provide a more coherent modeling of SRCs.

In this study, we will explore nuclear momentum anisotropy in intermediate-energy prolate uranium-uranium collisions using the isospin-dependent Boltzmann-Uehling-Uhlenbeck (IBUU) transport model. 
We will focus on several sensitive probes for studying geometric effects, such as elliptic flow, pion yields,  the $\pi^-/\pi^+$ ratio, etc.

\section{Model initialization \label{sec2}}

The Monte Carlo method-based IBUU transport model simulates the phase-space evolution of baryons and mesons during heavy-ion collisions, encompassing essential physical processes such as elastic and inelastic scattering, particle absorption, and decay \cite{Bertsch1988Phys.Rep.160.189233}.
The used IBUU model \cite{Yang2021J.Phys.GNucl.Part.Phys.48.105105, Yong2016Phys.Rev.C93.014602, Yang2018Phys.Rev.C98.014623,Guo2019Phys.Rev.C100.014617,Cheng2016Phys.Rev.C94.064621} has incorporates the Coulomb effect \cite{Yang2018Phys.Rev.C98.014623}, Pauli blocking, and medium effects on scattering cross sections \cite{Xu2011Phys.Rev.C84.064603}, etc.  

Considering that previous work has already provided us with some understanding of the geometric effects in U+U collisions \cite{Yang2024Phys.Lett.B848.138359}, this study will continue to use the stable, prolate nucleus $^{238}$U as the ideal research subject.
For the initialization of the nucleon density in coordinate space, a deformed Woods-Saxon form is adopted, expressed as
\begin{equation}
    \rho(r,\theta,\phi) = \frac{\rho_0}{1+e^{[r-R(\theta,\phi)]/a}},
\end{equation}
where the nuclear surface includes only the most relevant axial symmetric quadrupole deformation $R(\theta,\phi) = R_0(1+\beta_\text{r} Y_{20})$. 
The experimental value for the quadrupole deformation parameter is $\beta_\text{r}=0.29$ \cite{nudat}. 
In subsequent calculations, we also used spherically symmetric density $\beta_\text{r}=0$ for comparison.
The other parameters are taken from Ref.~\cite{Filip2009Phys.Rev.C80.054903}, with values $\rho_0=
0.168\,\mathrm{fm^{-3}}$, $a=0.54\,\mathrm{fm}$, and $R_0=6.81\,\mathrm{fm}$.

Since momentum and position are conjugate variables, we define a momentum quadrupole deformation parameter $\beta_\text{p}$ to describe the angular dependence of the Fermi surface, expressed as $P_\text{F}(\theta,\phi) = p_\text{f}(1+\beta_\text{p} Y_{20})$ with spherical Fermi momentum given by $p_\text{f} = \hslash (3 \pi^2 \rho_0 )^{1/3}$.
Therefore, for the initialization of momentum, we adopt a form similar to the coordinate distribution, in the Woods-Saxon shape, denoted as

\begin{equation}
    \label{EQ2}
    n(p,\theta,\phi) = \frac{1}{1+e^{[p-P_\text{F}(\theta,\phi)]/a_\text{p}}},
\end{equation}
where the tail diffuseness coefficient is set to $a_\text{p} = 0.01$ corresponding to 10\% of nucleons above the Fermi surface.
In principle, the given density and interactions correspond to an exact momentum distribution. 
Theoretically, both the local density approximation and the Fourier transform of the Kohn-Sham single-particle wave functions can be employed to calculate the momentum distribution \cite{Fan2022Phys.Lett.B834.137482}. 
However, experimental evidence is currently lacking to explore the correspondence between momentum and density, particularly in their geometric correlation.
The question is whether the momentum distribution of a nucleus that appears prolate in coordinate space is likewise prolate or, instead, oblate.

\begin{figure}[tb]
\includegraphics[width=7 cm]{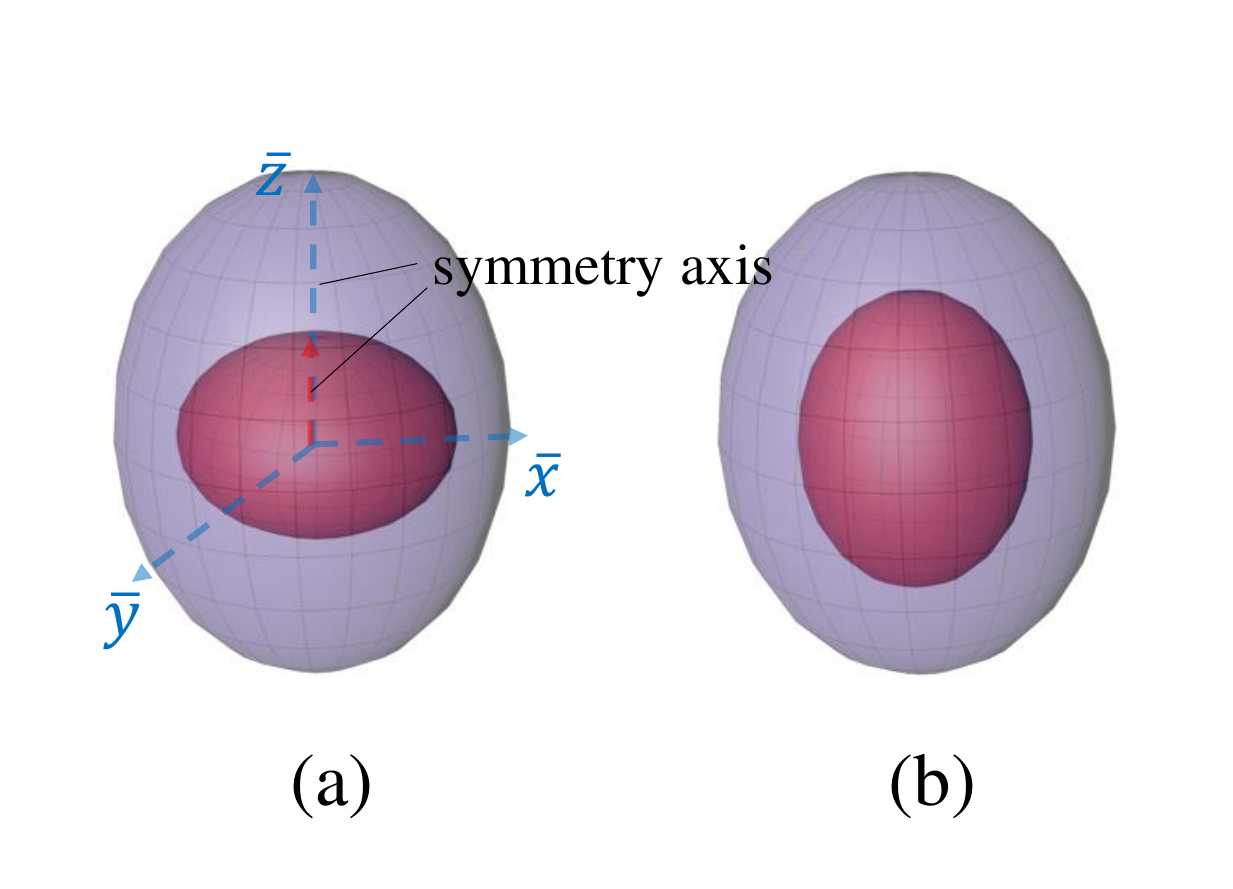}
\caption{\label{fig1} Diagram of the uranium density profile of nucleons in coordinate space (The outer blue ellipsoid) and momentum space (The inner red ellipsoid). 
The quadrupole deformation parameter in coordinate space is $\beta_\text{r} = 0.29$, and in momentum space is set to (a) $\beta_\text{p} = -0.29$ and (b) $\beta_\text{p} = 0.29$.
The $\overline{xyz}$ represents the intrinsic reference frame of the nucleus, with the $\overline{z}$-axis serving as the rotational symmetry axis.}
\end{figure}

To facilitate relevant discussions, the corresponding parameters are set to $\beta_\text{p}= -0.29$ and $\beta_\text{p}=0.29$, as shown in Fig.~\ref{fig1} (a) and (b), respectively.
In the figure, the outer blue ellipsoid depicts the nucleon density profile in coordinate space, while the inner red ellipsoid represents the nucleon distribution in momentum space.
Particularly, the relative orientation between the momentum ellipsoid and the coordinate ellipsoid is fixed in our simulations. 
When the rotational symmetry axis (intrinsic $\overline{z}$-axis) of the ellipsoid aligns with the transverse plane (perpendicular to the beam direction), the scenario is referred to as a body-body collision. 
In contrast, when the axis is parallel to the beam direction, it is termed a tip-tip collision.
It has been demonstrated in Ref.~\cite{Yang2024Phys.Lett.B848.138359, Yang2024.}, based on simulations of U+U reactions, that the collision orientation of prolate nuclei can be identified from the initial state through convolutional neural networks. 
This implies that selecting experimental data for tip-tip and body-body collision scenarios is feasible.

In this study, the isospin- and momentum-dependent mean-field single nucleon potential is used \cite{Yong2016Phys.Rev.C93.044610, Yong2017Phys.Rev.C96.044605}, i.e.,
\begin{equation}
\begin{aligned}
U(\rho, \delta, \vec{p}, \tau) &=  A_u(X) \frac{\rho_{\tau^{\prime}}}{\rho_0}+A_l(X) \frac{\rho_\tau}{\rho_0} \\
& +B\left(\frac{\rho}{\rho_0}\right)^\sigma\left(1-X \delta^2\right)-8 X \tau \frac{B}{\sigma+1} \frac{\rho^{\sigma-1}}{\rho_0^\sigma} \delta \rho_{\tau^{\prime}} \\
& +\frac{2 C_{\tau, \tau}}{\rho_0} \int d^3 \vec{p^{\prime}} \frac{f_\tau(\vec{r}, \vec{p^{\prime}})}{1+(\vec{p}-\vec{p^{\prime}})^2 / \Lambda^2} \\
& +\frac{2 C_{\tau, \tau^{\prime}}}{\rho_0} \int d^3 \overrightarrow{p^{\prime}} \frac{f_{\tau^{\prime}}(\vec{r}, \vec{p^{\prime}})}{1+(\vec{p}-\overrightarrow{p^{\prime}})^2 / \Lambda^2},
\end{aligned}
\end{equation}
where $\rho_0 = 0.168\, \text{fm}^{-3}$ is the empirical saturation density of nuclear matter and $\tau,\tau^{\prime} = 1/2(-1/2)$ is set for neutrons (protons).
The parameter values $A_u(X)$, $A_l(X)$, $B$, $C_{\tau, \tau}$, $C_{\tau, \tau^{\prime}}$, $\sigma$, and $\Lambda$ as well as the in-medium dependence on the scattering cross-section can be found in Ref.~\cite{Yong2016Phys.Rev.C93.044610}.
Particularly, the symmetry energy parameter is set to $X=1$.
Detail calculations on symmetry energy can be found in Ref.~\cite{Das2003Phys.Rev.C67.034611}.
In the simulation, the test particle method is used to stabilize the mean field, where the number of test particles $N$ is set to 50, meaning that an average of 50 point-particle events is used to simulate a real collision evolution.
More details on the particle-particle collisions can be found in Ref.~\cite{Yong2016Phys.Rev.C93.044610}.
It is worth mentioning that the employed interaction developed since 2016 \cite{Cheng2016Phys.Rev.C94.064621}, which fully matches experimental results at intermediate-energy, such as the S$\pi$RIT pion data in Sn + Sn systems \cite{Yong2021Phys.Rev.C104.014613}.

\section{Results and Discussion\label{sec3}}

Anisotropic flow is among the most sensitive observables for probing geometric effects in heavy-ion collisions \cite{Abdulhamid2024Nature635.6772, Yang2024Phys.Lett.B848.138359}. 
The second-order component of anisotropic flow, known as elliptic flow, exhibits a linear response to nuclear quadrupole deformation \cite{Abdulhamid2024Nature635.6772}, making it the primary observable for our investigation. 
It is defined as
\begin{equation}
v_2 = \left\langle \cos(2\phi) \right\rangle,
\end{equation}
where $\phi = \arccos{(p_x/p_\text{t})}$ is the azimuthal angle of transverse emission, and $p_\text{t} = \sqrt{p_x^2 + p_y^2}$ denotes the transverse momentum of emitted particles. 
Here, $\left\langle ... \right\rangle$ represents the average over all emitted particles.
The ultra-central body-body collision scenario maximally highlights the geometric effects in ellipsoidal nuclear reactions, and therefore our discussion of $v_2$ will primarily focus on this scenario.

\begin{figure}[tb]
\includegraphics[width=9 cm]{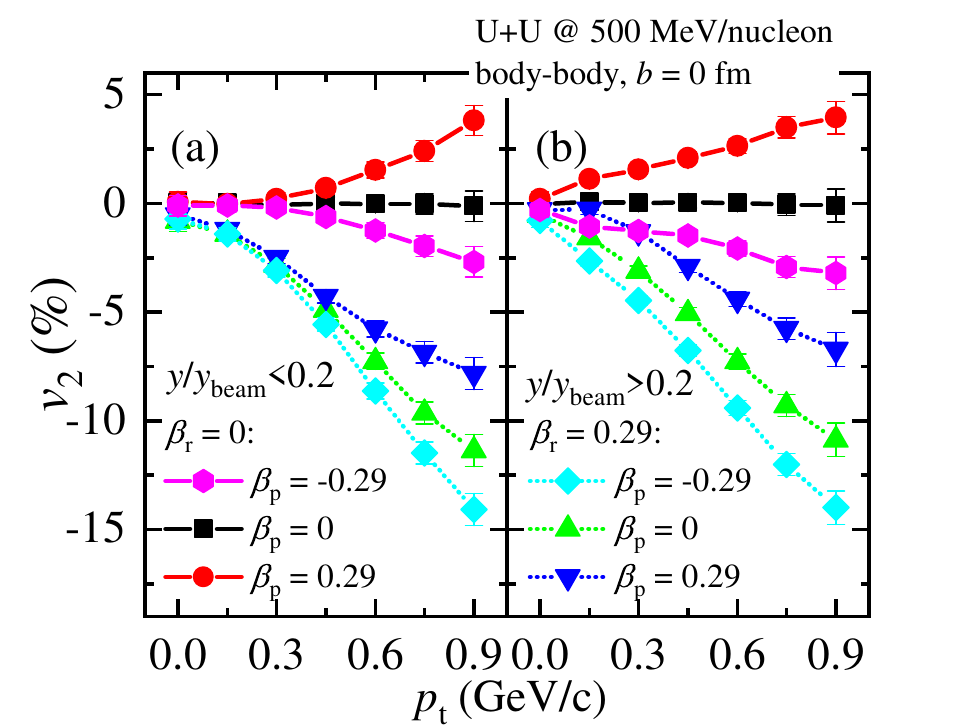}
\caption{\label{fig2} The elliptic flows $v_2$ as a function of transverse momentum $p_t(= \sqrt{p^2_x+p^2_y})$ at $(y/y_\text{beam})_\text{c.m.} \leq 0.2$ (a) and $(y/y_\text{beam})_\text{c.m.} \geq 0.2$ (b) for ultra-central body-body collisions of $^{238}\text{U}$ + $^{238}\text{U}$ at a beam energy of 500 MeV/nucleon with different quadrupole deformation parameters in coordinate (momentum) space.  
  }
\end{figure}

\begin{figure}[tb]
\includegraphics[width=9 cm]{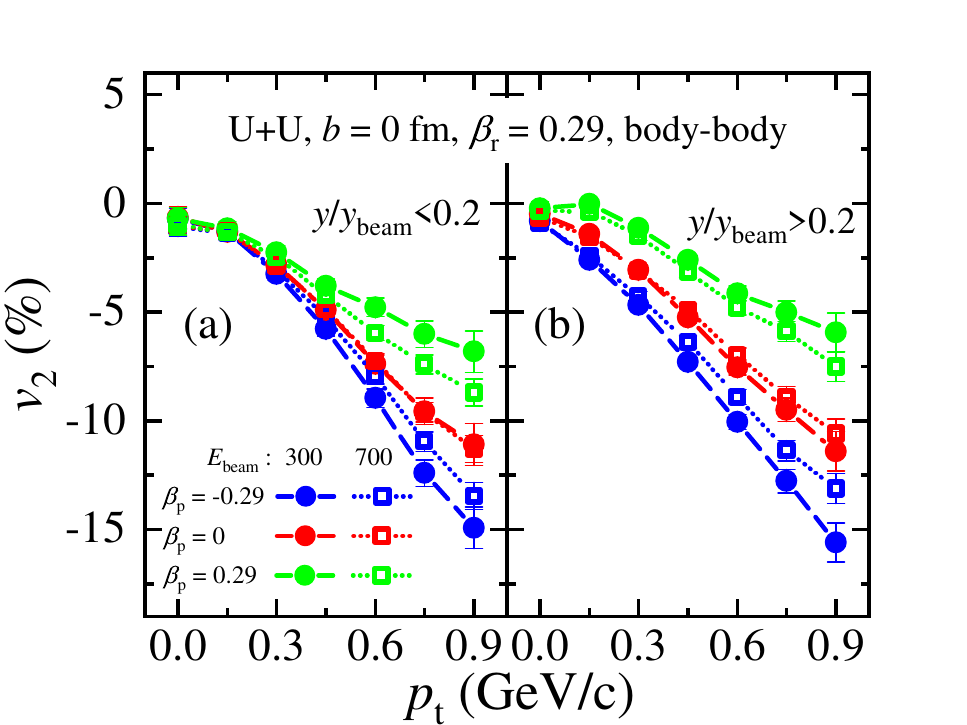}
\caption{\label{fig3} The elliptic flows $v_2$ as a function of transverse momentum $p_t$ at $(y/y_\text{beam})_\text{c.m.} \leq 0.2$ (a) and $(y/y_\text{beam})_\text{c.m.} \geq 0.2$ (b) for ultra-central body-body collisions of $^{238}\text{U}$ + $^{238}\text{U}$ at beam energies of 300 and 700 MeV/nucleon with different quadrupole deformation parameters in momentum space.  
  }
\end{figure}

The magnitude of the effect of momentum distribution deformation at 500 MeV/nucleon is first discussed, presented in Fig.~\ref{fig2}.
To compare nuclear deformation effects in the coordinate space, we utilize both spherically symmetric density ($\beta_\text{r} = 0$) and ellipsoidal density ($\beta_\text{r} = 0.29$) derived from laboratory measurements \cite{nudat}. 
For the momentum density, we set it to oblate ($\beta_\text{p} = -0.29$), spherical ($\beta_\text{p} = 0$), and prolate ($\beta_\text{p} = 0.29$), respectively, to explore the quadrupole deformation effect in momentum space.
As one can see, elliptic flow is absent when the deformation parameters in both the initial momentum and coordinate spaces are simultaneously zero.
Conversely, quadrupole deformations in either the initial coordinate space or momentum space will induce elliptic flow.

Aiming at the case of $\beta_\text{r} = 0$, oblate momentum density generates negative elliptical flow, while prolate momentum density produces positive elliptical flow, which becomes more pronounced at higher transverse momentum.
This is consistent with the definition (the symmetry axis of the ellipsoid is parallel to the $x$-axis of the transverse plane in the laboratory frame), as oblate momentum density implies a larger average $y$-component of the initial momentum, while prolate momentum density implies the opposite.
Further considering the inherent deformation in coordinate space $\beta_\text{r} = 0.29$, the elliptical flow intensity becomes more pronounced.
It is worth noting that the prolate nucleon density distribution leads to negative elliptical flow, which is the opposite of the situation with momentum.
This is because nucleons are more easily compressed along the $y$-direction in a body-body collision.
Besides, comparing the cases with and without coordinate space deformation, it is evident that the elliptical flow shift caused by momentum distribution deformation is almost identical. 
We conclude that momentum and coordinate deformation effects on elliptical flow are linearly additive.
The distribution in coordinate space still dominates the elliptical flow, while consistent signs of momentum and coordinate deformation parameters weaken the elliptical flow, and opposite signs enhance it.

Furthermore, comparing panel (a) and panel (b) reveals that the beam/target rapidity region ($(y/y_\text{beam})_\text{c.m.} \geq 0.2$) is more sensitive to momentum anisotropy. 
In the mid-rapidity region ($(y/y_\text{beam})_\text{c.m.} \leq 0.2$), particles with low transverse momentum show less sensitivity to momentum distribution deformation.
This can be attributed to the greater influence of high-momentum nucleons in the initial state (the nucleons in the initial momentum ellipsoidal outer layers), which tend to produce high-rapidity particles after the reaction.
From another perspective, when measuring $\beta_\text{r}$ alone, greater attention should be focused on the low transverse momentum region in mid-rapidity.

The behavior of momentum anisotropy at different beam energies warrants further investigation.
Figure \ref{fig3} shows the effect of momentum distribution deformation on elliptical flow at beam energies of 300 MeV/nucleon and 700 MeV/nucleon.
By examining the red curves, it can be observed that whether at mid- or the beam/target rapidity region, when the momentum is spherical, the beam energy has little effect on the magnitude of the elliptic flow.
This suggests that in the present reaction system, the variation in energy has a modest impact on the geometric effects in coordinate space.
When deformation in momentum space is included in the calculation, it is noticeable that as the energy increases, the elliptic flow curve gradually approaches the curve corresponding to spherical momentum density.
This indicates that the increase in energy weakens the effects of momentum anisotropy.
As an expansion, this also implies that extracting momentum density deformation at ultra-relativistic energies is impractical. 
In this instance, the anisotropic flow becomes more sensitive to the distribution in coordinate space.
In contrast, incorporating momentum density deformation may be fruitful when considering low-energy fusion reactions, such as in the exploration of element 119, 120 syntheses.

\begin{figure}[tb]
\includegraphics[width=8 cm]{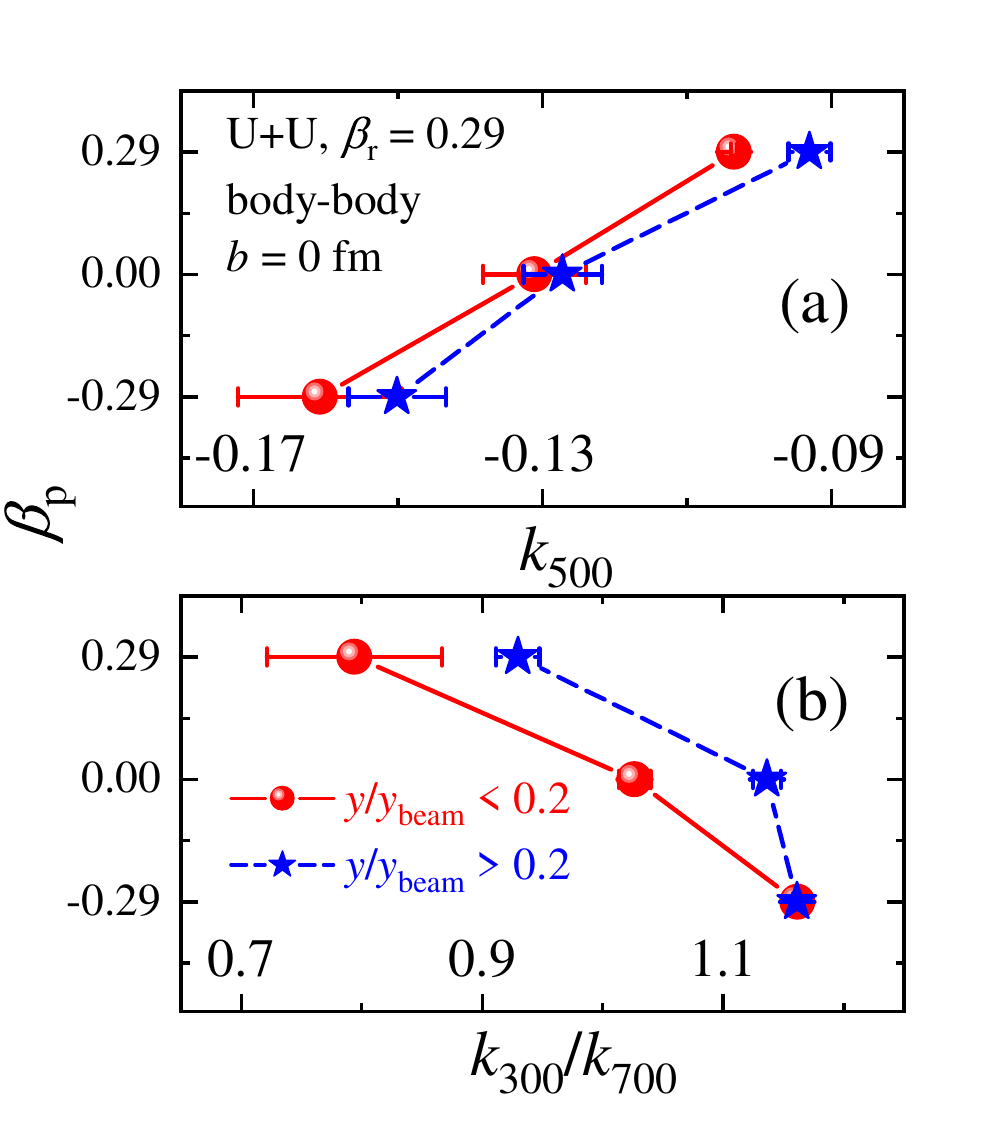}
\caption{\label{fig4} (a): The response relation between the momentum deformation parameter $\beta_\text{p}$ and the slope $k_{500}$ of $v_2(p_\text{t})$  (from $ p_\text{t} = 0.15$  to  0.6  GeV/c) for body-body collisions at a beam energy of 500 MeV/nucleon.
(b): The $\beta_\text{p}$ corresponding to the slope ratio $k_{300}/k_{700}$ of $v_2(p_\text{t})$ (from $ p_t = 0.15 $ to $ 0.6 $ GeV/c) as the beam energy varies from 300 MeV/nucleon to 700 MeV/nucleon. 
  }
\end{figure}

According to the previous discussion, it is noted that the momentum deformation parameter $\beta_\text{p}$ can be experimentally extracted through the slope of the  $v_2$-$p_t$ relationship.
The response relation between $\beta_\text{p}$ and the slope $k_{500}$ of $v_2(p_\text{t})$  (from $ p_\text{t}$ = 0.15  to  0.6  GeV/c) for body-body collisions at a beam energy of 500 MeV/nucleon is plotted in Fig.~\ref{fig4}(a), which is obtained through the least squares method.
In this situation, the slopes across different intervals demonstrates a robust linear relationship with momentum anisotropy, indicating the distinguishability of $\beta_\text{p}$.
To further eliminate potential systematic biases in the response relationship, we calculate the slope ratio $k_{300}/k_{700}$ of the two systems at beam energies of 300 MeV/nucleon and 700 MeV/nucleon as a more reliable experimental reference, which is shown in Fig.~\ref{fig4}(b).
Clearly, the increased slope ratio corresponds to a more oblate momentum distribution.
In the mid-rapidity region, the slope ratio is more sensitive and can be used to distinguish momentum distribution from prolate to oblate. 
However, in the beam/target rapidity region, the distribution differences between spherical and oblate shapes are not as distinct.

\begin{figure}[tb]
\includegraphics[width=8.5 cm]{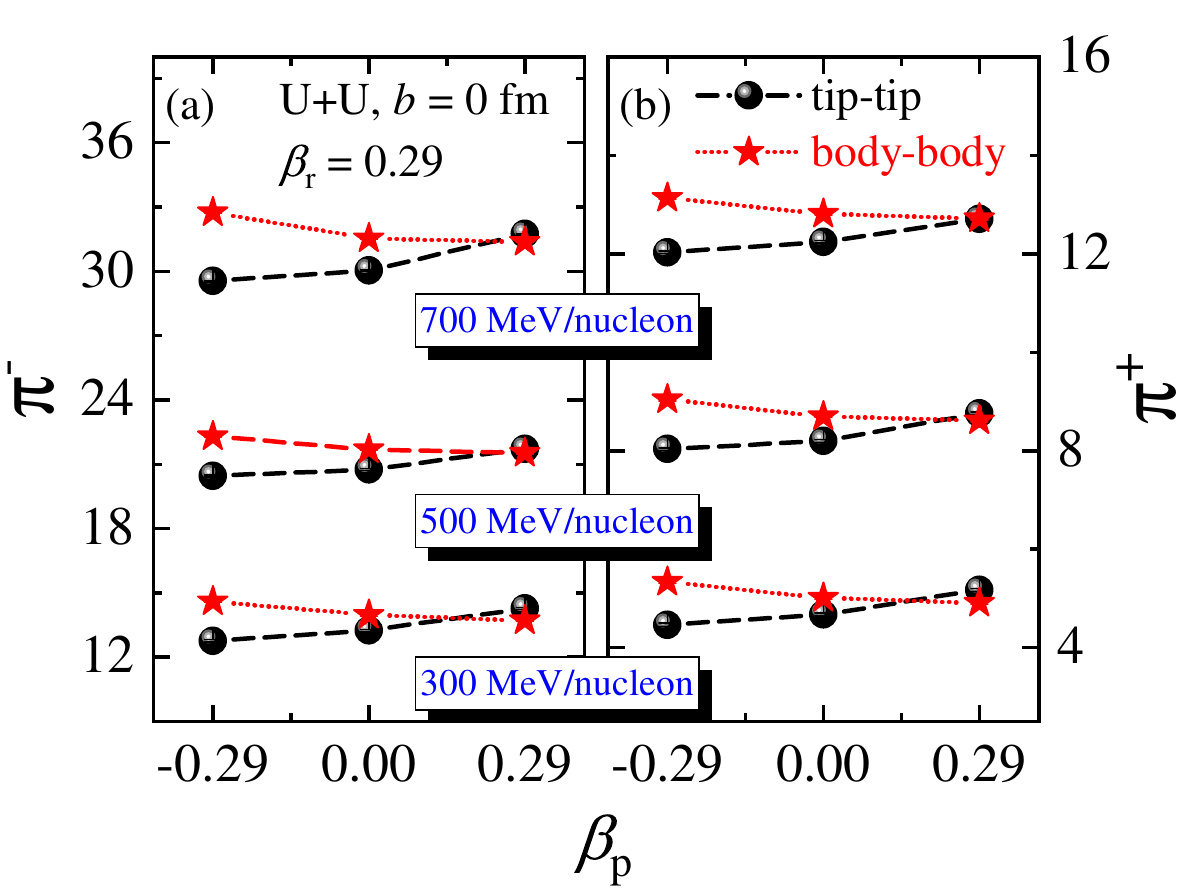}
\caption{\label{fig5} The multiplicities of (a) $\pi^-$ and (b) $\pi^+$ as a function of the momentum space deformation parameter for ultra-central body-body and tip-tip collisions of $^{238}\text{U}$ + $^{238}\text{U}$ at beam energies of 300, 500, and 700 MeV/nucleon.  
  }
\end{figure}

The pion meson is another key observable, influenced by the reaction orientation and the initial momentum.
The $\pi^-$ and $\pi^+$ multiplicities under different orientations and momentum deformation parameters are calculated and presented in Fig.~\ref{fig5}.
Across all calculated energies, the trend in meson production remains consistent: 
in tip-tip collisions, meson yields increase with a higher momentum deformation parameter, while in body-body collisions, yields decrease under the same conditions. 
This trend arises because, in tip-tip collisions, increasing the momentum deformation parameter enhances the projection of nucleon momentum along the beam direction, thereby increasing the reaction pressure and meson production. 
Conversely, in body-body collisions, the prolate momentum ellipsoid produces a smaller momentum projection along the $z$-axis. 

Additionally, meson yields in the two orientations converge only when the momentum deformation parameter approaches or even exceeds the coordinate deformation parameter. 
In other cases, body-body collisions consistently produce more mesons.
According to our simulations, the reaction duration in tip-tip collisions is approximately 15\% longer than in body-body collisions, which may lead to more nucleon escape, thereby reducing meson production.

\begin{figure}[tb]
\includegraphics[width=8.5 cm]{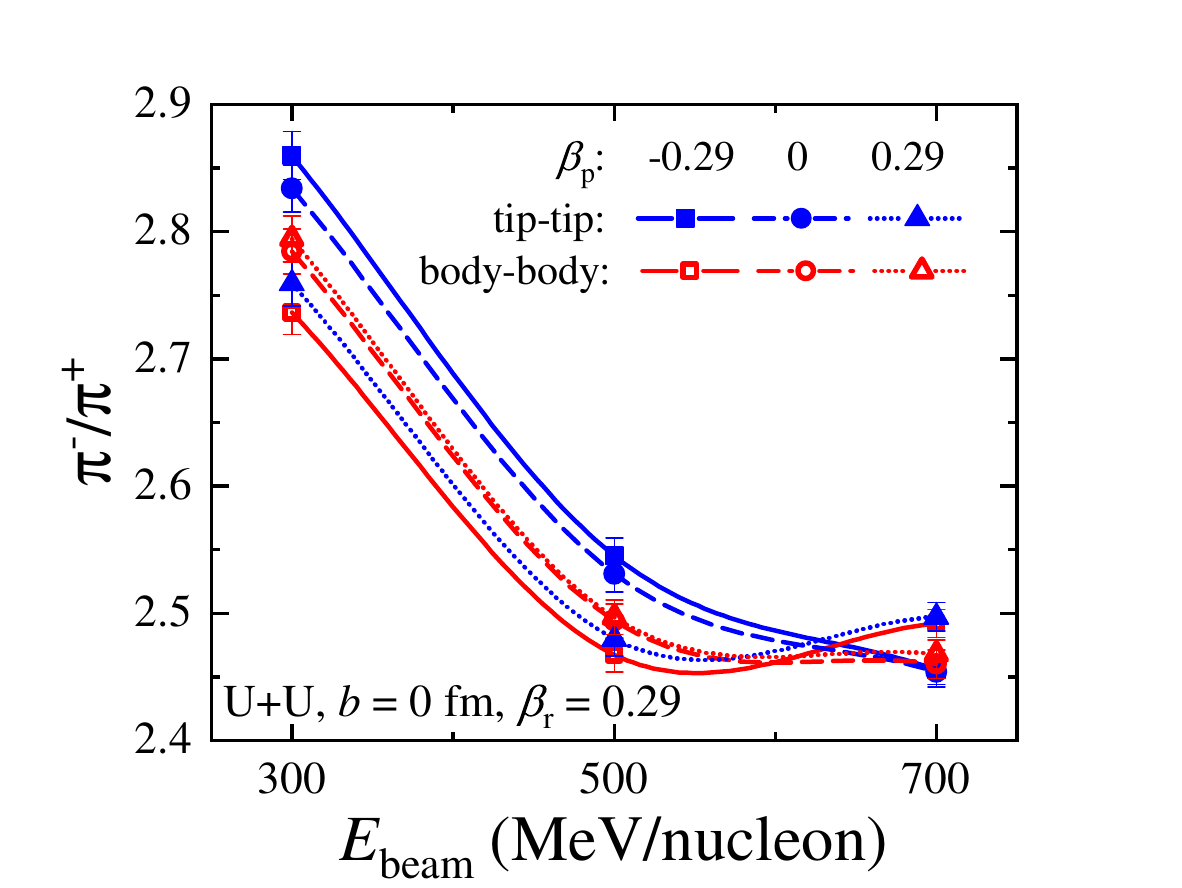}
\caption{\label{fig6}The $\pi^-/\pi^+$ ratio as a function of the beam energy for ultra-central body-body and tip-tip collisions of $^{238}\text{U}$ + $^{238}\text{U}$.  
  }
\end{figure}

Turning to the $\pi^-/\pi^+$ ratio, its variation with energy for different orientations is shown in Fig.~\ref{fig6}.
As the beam energy increases from 300 to 700 MeV/nucleon, the $\pi^-/\pi^+$ ratio basically decreases, which is consistent with the calculations in Ref.~\cite{Cozma2018Eur.Phys.J.A54.}.
From 300 to 500 MeV/nucleon, the $\pi^-/\pi^+$ ratio maintains a consistent trend across different orientations and momentum deformation parameters.
This energy range is relatively broad, which is advantageous for experimental measurements.
Particularly, within this energy range, we observe that the differences in reaction results are most pronounced for oblate momentum density in the two orientations. 
This is understandable: in tip-tip collisions, the oblate deformation increases momentum in the transverse plane, enhancing the likelihood of particles being squeezed out under the influence of the Coulomb force. 
This, in turn, reduces proton participation in the reaction, lowers the yield of $\pi^+$, and consequently increases the $\pi^-/\pi^+$ ratio.
For body-body collisions, the situation is the opposite.
It can be observed that as the energy increases, the curves converge to some extent, which also supports the notion that the anisotropy of momentum plays a smaller role as the energy increases.
Surprisingly, at 700 MeV/nucleon, the situation changes: the $\pi^-/\pi^+$ ratio increases for cases with larger $z$-axis momentum projections (oblate momentum in body-body and prolate momentum in tip-tip) compared to 500 MeV/nucleon.
This is possibly because larger $z$-axis momentum projections tend to produce more $\pi^-$ mesons, thereby preventing the ratio from decreasing.
The anomalous situation warrants further investigation in the future.

\begin{figure}[tb]
\includegraphics[width=8.5 cm]{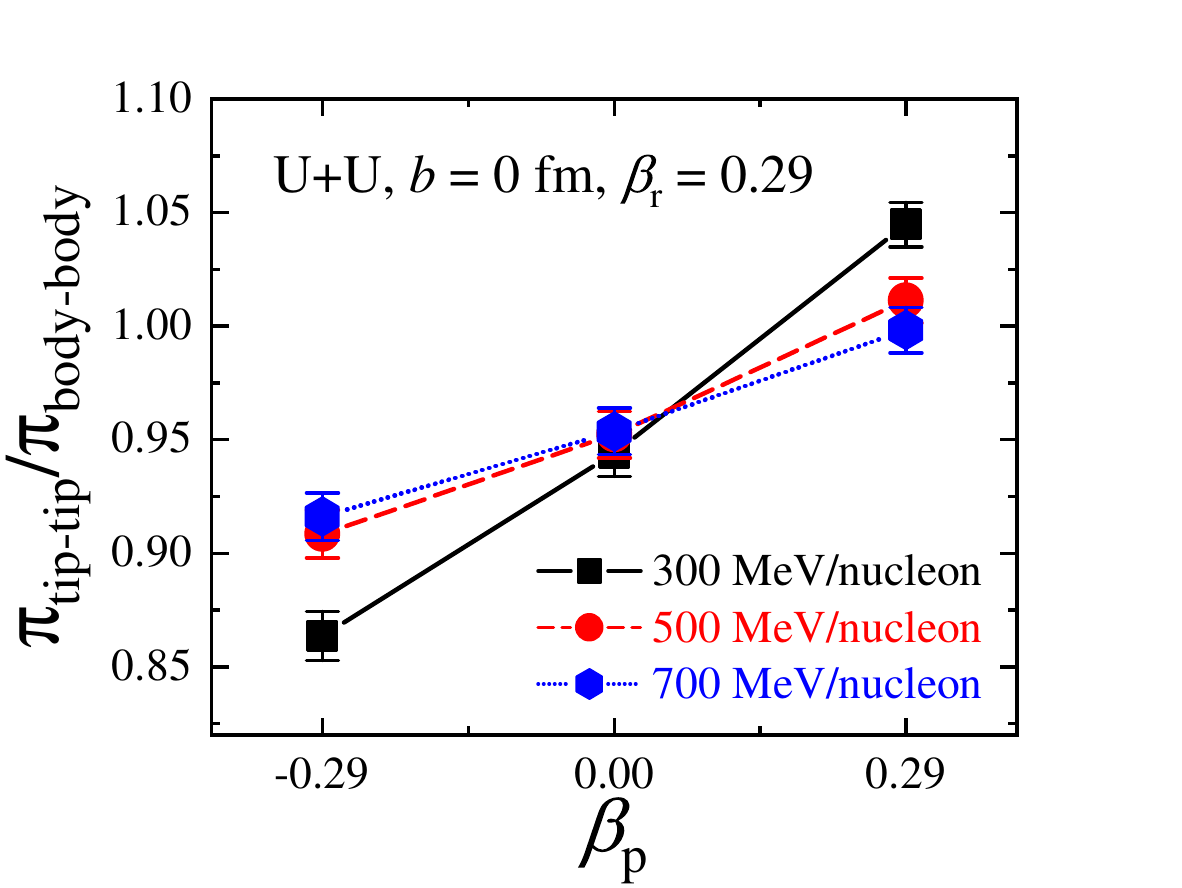}
\caption{\label{fig7} The ratio of charged pion yields between ultra-central tip-tip and body-body collisions of $^{238}\text{U} + ^{238}\text{U}$ as a function of the momentum deformation parameter, at beam energies of 300, 500, and 700 MeV/nucleon.
  }
\end{figure}

The ratio of charged pions in tip-tip and body-body collisions ($\pi_\text{tip-tip}/\pi_\text{body-body}$) has also been further investigated, shown in Fig.~\ref{fig7}.
It is observed that the ratio increases linearly as the momentum deformation parameter $\beta_\text{p}$ increases.
Furthermore, the slope for $\pi_\text{tip-tip}/\pi_\text{body-body}$ to $\beta_\text{p}$ decreases as the energy increases. 
These conclusions align with our previous discussion and further support that capturing momentum anisotropy is only feasible at medium to low energies.
In light of the discussions regarding Figs.~\ref{fig6} and \ref{fig7}, we assert that extracting momentum deformation parameters from experiments conducted at 300-500 MeV/nucleon is reliable.

The above conclusions require determining the collision orientation. 
To facilitate experimental implementation, we employ non-polarized collisions for further validation. 
In this setup, the projectile and target are each subjected to random-angle 3D Euler rotations \cite{Yang2024Phys.Lett.B848.138359}, with the impact parameter set to $b \in [0,1]$ fm and the beam energy fixed at 500 MeV/nucleon. 
Aiming at the events with uncertain orientation, experimentalists can measure the elliptic flow on an event-by-event basis and calculate its mean square,  $\left\langle v_2^2 \right\rangle$, to compare with theoretical predictions. 
The theoretical values for the mean square elliptic flow are provided in Table~\ref{tab1}.

\begin{table}[h]
\centering
\caption{\label{tab1} Mean square elliptic flow $\left\langle v_2^2 \right\rangle$ ($\times 10^{-5}$) in non-polarized ultra-central ($b<1$ fm) $^{238}\text{U}$ + $^{238}\text{U}$ collisions at beam energies of 500 MeV/nucleon. The $\delta\left\langle v_2^2 \right\rangle$ is the statistical error.} 
\renewcommand{\arraystretch}{1.2}
\begin{tabular}{l|llll}
\hline\hline
     & \multicolumn{2}{c}{$\beta_\text{r}=0.29$~~~~~~~~} & \multicolumn{2}{c}{$\beta_\text{r}=0$~~~~~~~~~~~~}                       \\ 
     & $\left\langle v_2^2 \right\rangle$~~~            & \multicolumn{1}{l}{$\delta\left\langle v_2^2 \right\rangle$ }          & $\left\langle v_2^2 \right\rangle$~~~        & \multicolumn{1}{l}{$\delta\left\langle v_2^2 \right\rangle$}         \\ \hline
~~$\beta_\text{p}= -0.29$~~ & 25.071   & 0.237   & 2.636 & 0.134               \\
~~$\beta_\text{p}= 0$~~ & 13.588   & 0.194  & 1.391 & 0.094               \\
~~$\beta_\text{p}= 0.29$~~ & 6.120   & 0.253   & 3.293 & 0.154               \\
 \hline\hline
\end{tabular}
\end{table}

The change in $\left\langle v_2^2 \right\rangle$  caused by quadrupole deformed momentum is quite significant. 
Along the symmetry axis in coordinate deformation, the elongated momentum density weakens $\left\langle v_2^2 \right\rangle$, while the compressed one enhances $\left\langle v_2^2 \right\rangle$.
When performing initial-state orientation recognition, enhancing anisotropy implies an improvement in recognition accuracy \cite{Yang2024Phys.Lett.B848.138359}, which suggests that an oblate momentum density would favor orientation recognition.
For a spherical nucleus, the appearance of any momentum anisotropy will logically lead to an increase in $\left\langle v_2^2 \right\rangle$.
These conclusions will help us more thoroughly understand nuclei's intrinsic properties, especially the impact of nuclear interactions on momentum space.

\section{summary \label{sec:5}}

Based on the intermediate-energy ultra-central deformed uranium-uranium collision simulated using the IBUU transport model with a momentum-dependent mean-field single nucleon potential, we investigate the impact of momentum quadrupole deformation on geometric effects.
To this end, we define a momentum quadrupole deformation parameter to establish an ellipsoidal Fermi surface, aligning its symmetry axis with the one in coordinate space.

We examine the impact of momentum density deformation on elliptic flow in body-body collisions.
It is found that the oblate momentum distribution enhances the elliptic flow generated by prolate deformation in coordinate space, while prolate momentum distribution weakens the elliptic flow.
The influence of momentum anisotropy on geometric effects is not significant in the mid-rapidity, low transverse momentum region, which is favorable for measuring nuclear deformation with avoiding the influence of momentum.
To provide clearer guidance for experiments, we calculate the slope $k_{500}$ of elliptic flow as a function of transverse momentum at 500 MeV/nucleon and further compute the ratio of the two systems $k_{300}/k_{700}$ to eliminate systematic errors.
Non-polarized U+U reactions are also explored, and significant effects are observed on the squared mean elliptic flow as well.

On the other hand, momentum anisotropy also leads to differences in the initial momentum mean projection along the beam direction, and the larger momentum projection produce more pion mesons are produced.
In our simulations, the  $\pi^-/\pi^+$ ratio maintains the same trend of variation at 300-500 MeV/nucleon. 
During this phase, oblate momentum deformation is most sensitive to orientations.
The ratio of charged pions in tip-tip and body-body collisions ($\pi_\text{tip-tip}/\pi_\text{body-body}$) is further investigated, which increases linearly as the momentum deformation parameter $\beta_\text{p}$ increases.

For all observables, we notice that the lower the energy, the more pronounced the momentum deformation effects. 
This suggests that extracting geometric properties is more reliable at relativistic energies, as it largely mitigates the impact of momentum. 
In contrast, when simulating reactions at lower energies, such as in the fusion synthesis of element 119, 120, considering the anisotropy of the momentum distribution can enhance reliability.

\section{Acknowledgements}
We acknowledge helpful discussions with Profs.~ Gaochan Yong and Haozhao Liang.
This work is supported by the National Natural Science Foundation of China under Grants Nos. 12005175, 12375126,
the Fundamental Research Funds for the Central Universities under Grant No.~SWU-KT24005,
the JSPS Grant-in-Aid for Scientific Research (S) under Grant No. 20H05648, 
and the RIKEN Projects: r-EMU, RiNA-NET, and the INT Program INT-23-1a and Institute for Nuclear Theory.

\bibliographystyle{apsrev4-1}
\bibliography{Ref}

\end{document}